\documentclass[
aps,
prx,
reprint,
amsmath,
amssymb,
superscriptaddress,
floatfix]{revtex4-1}

\usepackage{xr}

\usepackage[pdftex]{graphicx}
\usepackage{amsfonts}
\usepackage{amsmath}
\usepackage{amssymb}
\usepackage{graphicx}
\usepackage{color}
\usepackage{amsmath}
\usepackage{float}
\usepackage{hyperref}
\usepackage{gensymb}
\usepackage{multibib}
\usepackage{upgreek}

\makeindex

\begin{document}
\title{Hyperuniform monocrystalline structures by spinodal solid-state dewetting}

\author{Marco Salvalaglio} \email{marco.salvalaglio@tu-dresden.de} 
\affiliation{Institute  of Scientific Computing,  TU  Dresden,  01062  Dresden,  Germany}
\affiliation{Dresden Center for Computational Materials Science (DCMS), TU  Dresden,  01062  Dresden,  Germany}
\author{Mohammed Bouabdellaoui}
\affiliation{Aix Marseille Univ, Universit\'e de Toulon, CNRS, IM2NP, Marseille, France}
%
%
\author{Monica Bollani}
\email{monica.bollani@ifn.cnr.it} 
\affiliation{Istituto di Fotonica e Nanotecnologie-Consiglio Nazionale delle Ricerche, Laboratory for Nanostructure Epitaxy and Spintronics on Silicon, Via Anzani 42, 22100 Como, Italy.}
\author{Abdennacer Benali}
\affiliation{Aix Marseille Univ, Universit\'e de Toulon, CNRS, IM2NP, Marseille, France}
%
%
%
\author{Luc Favre}
\affiliation{Aix Marseille Univ, Universit\'e de Toulon, CNRS, IM2NP, Marseille, France}
\author{Jean-Benoit Claude}
\affiliation{Aix Marseille Univ, CNRS, Centrale Marseille, Institut Fresnel, 13013 Marseille, France}
\author{Jerome Wenger}
\affiliation{Aix Marseille Univ, CNRS, Centrale Marseille, Institut Fresnel, 13013 Marseille, France}
%
%
%
%
%
\author{Pietro de Anna}
\affiliation{Institut des Sciences de la Terre, University of Lausanne, Lausanne 1015, Switzerland}

\author{Francesca Intonti}
\affiliation{LENS, University of Florence, Sesto Fiorentino, 50019, Italy}

%
\author{Axel Voigt} 
\affiliation{Institute  of Scientific Computing, TU  Dresden,  01062  Dresden,  Germany}
\affiliation{Dresden Center for Computational Materials Science (DCMS), TU  Dresden,  01062  Dresden,  Germany}
\author{Marco Abbarchi}
\email{marco.abbarchi@im2np.fr}
\affiliation{Aix Marseille Univ, Universit\'e de Toulon, CNRS, IM2NP, Marseille, France}
 
\begin{abstract}
Materials featuring anomalous suppression of density fluctuations over large length scales are emerging systems known as disordered hyperuniform. The underlying hidden order renders them appealing for several applications, such as light management and topologically protected electronic states. These applications require scalable fabrication, which is hard to achieve with available top-down approaches. Theoretically, it is known that spinodal decomposition can lead to disordered hyperuniform architectures. Spontaneous formation of stable patterns could thus be a viable path for the bottom-up fabrication of these materials. Here we show that mono-crystalline semiconductor-based structures, in particular Si$_{1-x}$Ge$_{x}$ layers deposited on silicon-on-insulator substrates, can undergo spinodal solid-state dewetting featuring correlated disorder with an effective hyperuniform character. Nano- to micro-metric sized structures targeting specific morphologies and hyperuniform character can be obtained, proving the generality of the approach and paving the way for technological applications of disordered hyperuniform metamaterials. Phase-field simulations explain the underlying non-linear dynamics and the physical origin of the emerging patterns. 
%
%
\end{abstract}
\maketitle

Disordered hyperuniform materials are an emerging class of systems featuring anomalous suppression of density fluctuations on large length scales ~\cite{Torquato2003,TORQUATO20181,KLA2019,TOR2018,PIE2018}. While not presenting any Bragg peak in diffraction (as a liquid), they have strongly suppressed density fluctuations at long distances (as an ordered crystal). The underlying hidden order results in exotic phenomena, such as topologically protected electronic states~\cite{MIT2018}, glassy electronic quantum state transitions~\cite{GER2019}, Anderson localization of light~\cite{FRO2017}, polarization selectivity~\cite{ZHO2016}, lasing~\cite{DEG2016} and a full photonic band-gap for light propagation~\cite{FLO2009,MAN2013a,MUL2014,FRO2016}. 

Patterns exhibiting correlated disorder are ubiquitous in nature and are typical of many phenomena ruled by far-from-equilibrium processes~\cite{CRO1993,BAL1999}. Prominent examples are morphogenesis in biological systems~\cite{TUR1990}, thin-layer wrinkling~\cite{Becker2003,STO2015}, and phase separation~\cite{CAH1961}. In these systems the presence of interactions and the underlying non-linear dynamics, lead to the formation of complex patterns eventually featuring correlated disorder~\cite{HER1998}. Such patterns can emerge from phenomena involving long-range interactions and have been reported in thin films of polymers~\cite{XIE1998,HIG2000} and liquid metals~\cite{HER1998}. 
Moreover, phase separation by spinodal decomposition has been recently proposed as a possible bottom-up process for producing hyperuniform materials~\cite{Ma2017,ma2020}.

Liquid thin films of metals and polymers can break and dewet through the amplification of uniformly distributed surface undulations~\cite{HER1998,XIE1998,HIG2000}. Owing to the similarities of the final morphologies to those observed in phase separation via spinodal decomposition, this process is commonly termed spinodal dewetting. \textit{De facto}, this regime has remained mostly inaccessible in the technologically relevant case of semiconductors and, more generally, in monocrystalline systems. Indeed, in thin crystalline films~\cite{XIE1998,Thompson2012} the melting temperature is typically too high, rendering impossible to access the same spinodal dewetting dynamics. Instead, they undergo heterogeneous nucleation of holes, rim formation and retraction, followed by fingers and islands formation well-below their melting temperature. Owing to the anisotropic surface diffusion, the islands are aligned along the crystallographic directions and are formed at a typical distance set by the initial layer thickness~\cite{Naffouti2016}.

\begin{figure*}[ht!]
    \centering
   \includegraphics[width=0.95\textwidth]{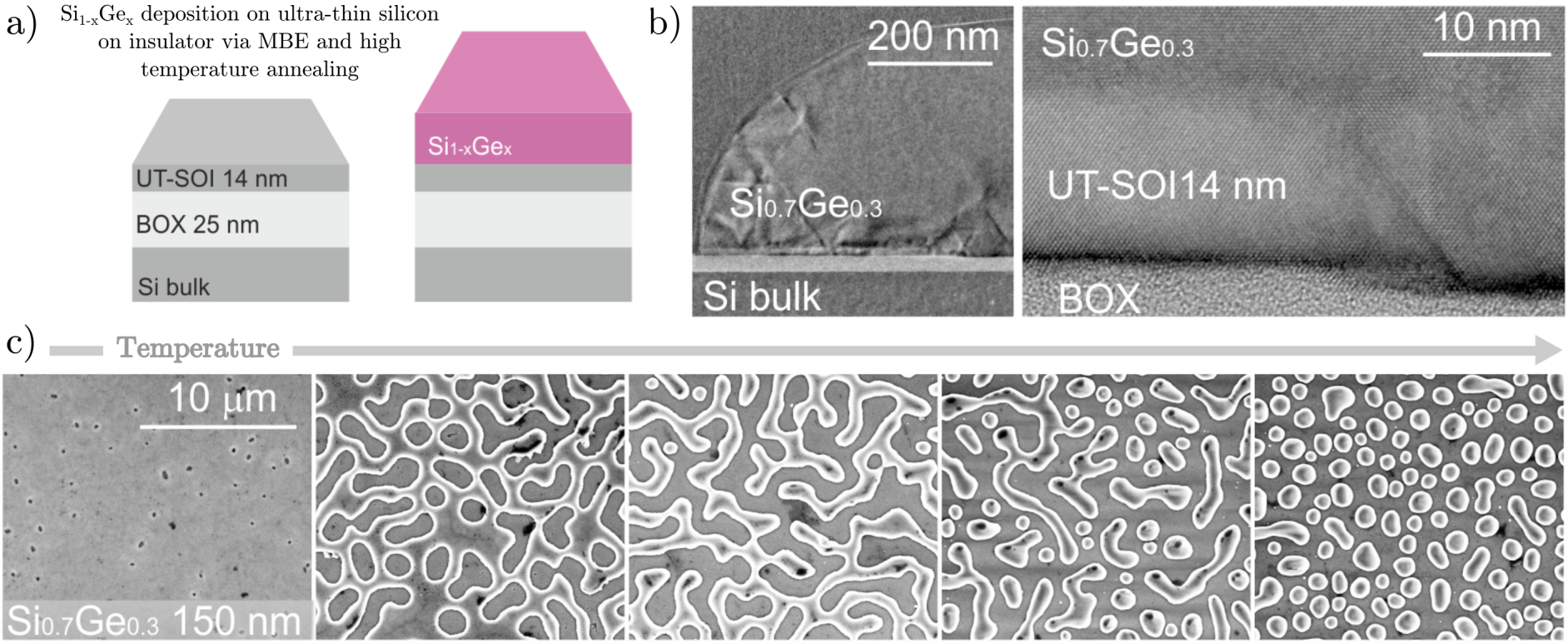} 
   \caption{
   \textit{Spinodal solid-state dewetting}. a) Scheme of the samples: ultra-thin silicon on insulator substrates (14~nm thick Si on 25~nm thick SiO$_{2}$ BOX) followed by epitaxial deposition of Si$_{1-x}$Ge$_{x}$ alloys and high-temperature annealing in a MBE. b) Left panel: transmission electron micrograph (TEM) of a dewetted island. Right panel: high-resolution TEM highlighting the interface between BOX, some pristine UT-SOI and Si$_{0.7}$Ge$_{0.3}$. c) Scanning electron micrographs (SEM) of 150~nm Si$_{0.7}$Ge$_{0.3}$ on UT-SOI after 4 hours annealing at a nominal temperature of 800$\celsius$ (measured with a pyrometer at the center of the sample). Left to right: morphologies observed from the edge to the center of the sample, respectively taken at 150 $\mu$m, 500 $\mu$m, 920 $\mu$m, 1800 $\mu$m, and 7500 $\mu$m from the sample edge (the latter being unchanged up to the center and starting at $\sim$3~mm from the edge). They account for a temperature gradient of about 50$\celsius$ and allow to monitor a smooth change in a single experiment.}
    \label{fig:figure1}
\end{figure*}

Here we report on the annealing of semiconductor, monocrystalline, thin films showing that: 1) they can undergo spinodal solid-state dewetting, as in Refs.~\onlinecite{XIE1998,HER1998,Thompson2012}, and 2) the process leads to morphologies featuring different topology (connected or disconnected structures), size (from 0.1~$\mu$m up to 10~$\mu$m) and a strong hyperuniform character. These properties are shown for Si$_{1-x}$Ge$_{x}$ layers  (with $x$ = 0.3 to 1) deposited on ultra-thin silicon on insulator (UT-SOI). Three-dimensional phase-field simulations map the experimental observations. Exploiting Minkowski functionals, the arrangement of the resulting structures is shown to deviate from random patterns. Their effective hyperuniformity is assessed through metrics derived a spectral-density analysis~\cite{Torquato2016,Ma2017}.

The experimental system is illustrated in Fig.~\ref{fig:figure1}(a): Si$_{1-x}$Ge$_{x}$-based thin films ($x= 0.3-1$, thickness $5-2000$~nm), are deposited in a molecular beam epitaxy reactor (MBE) on 14~nm thick monocrystalline UT-SOI (on 25~nm thick buried oxide, BOX) previously diced in 2~cm $\times$ 2~cm samples. Here, they undergo high-temperature annealing (400-800$\celsius$ for 0.5-6 hours) under ultra-high vacuum  ($\sim$10$^{-10}$~Torr). 

A representative set of structures obtained by dewetting is reported in Fig.~\ref{fig:figure1}(b)-(c): exploiting a temperature gradient from the sample edge towards its center (roughly of about 50$\celsius$) we observe a continuous morphological change of the sample surface from holes to connected (noodles-like-) structures, and finally, isolated islands. A similar sequence of structures can be  obtained by focusing on the same region in the sample after annealing at different temperature (see other experiments in the following). 
When using thick Si$_{0.7}$Ge$_{0.3}$ layers (e.g., $>$200~nm), after dewetting, we observe structures with threading dislocations propagating along the (111) plane (Fig.~\ref{fig:figure1}(b)), as expected in crystalline bilayer systems with lattice mismatch. 

Surface corrugations~\cite{FRI2014} and island formation~\cite{Aqua2013,SHK2016} in strained Si$_{1-x}$Ge$_{x}$ films on Si (with partial or absent plastic relaxation) are commonly described by the Asaro-Tiller-Grinfeld (ATG) instability  \cite{AsaroMT1972,GrinfeldJNS1993,SrolovitzAMETAL1989}, i.e. as the growth of 3D structures relaxing in-plane strains with a periodicity determined by the balance between elastic and surface energy.
On bulk Si, trenches between islands form and deepen within the substrate while mass can flow across them. No 
dewetting occurs while coarsening takes place on a bulk substrate. The key difference in the process illustrated in Fig.~\ref{fig:figure1} is the use of UT-SOI: the ATG instability provides the driving force for digging into the UT-SOI, suddenly uncovers the BOX underneath and there initiates solid-state dewetting. 

This dewetting instability is simultaneous across the sample (in contrast with conventional solid-state dewetting~\cite{Ye2011,Thompson2012,Naffouti2016,Ye2018,BOL19}) and we term the process illustrated in Fig.~\ref{fig:figure1} \textit{spinodal solid state dewetting} due to the morphology of the resulting disordered structures: 
in spite of the different forces at play and material-transport mechanisms, they resemble those obtained in thin films of polymers and liquid metals~\cite{Becker2003,STO2015,HER1998,HER2012,Galinski2014,Mantz_2008}.

\begin{figure}[t!]
    \centering
    \includegraphics[width=\linewidth]{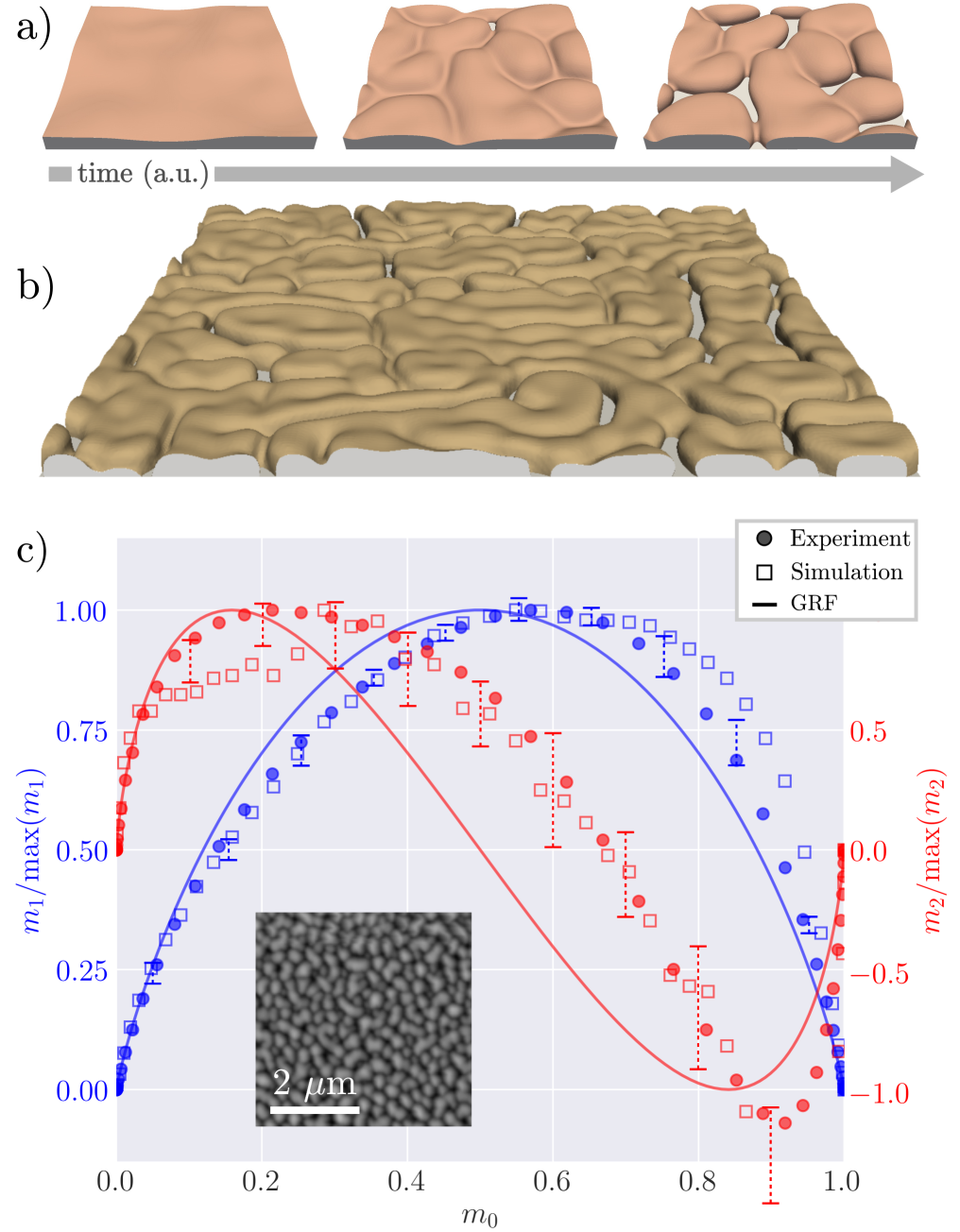}   
    \caption{\textit{Simulations and pattern analysis}. (a) Breakup of a perturbed film ($h=0.25\ell$, $L=4\ell$, details on $\ell$ in the text). (b) Simulated sample for pattern analysis ($h=0.5\ell$, $L=20\ell$). (c) Analysis by Minkowski functionals ($m_1(\bar{\rho})$ and $m_2(\bar{\rho})$ vs $m_0(\bar{\rho})$) for the pattern of (b) (open squares) and experimental data (filled dots), corresponding to 25~nm Ge deposited on UT-SOI and annealed at 470$\celsius$ for 45 minutes. The inset shows a $1/10\times1/10$ portion of the full image (reported in the Supplemental Material, S.M.). The error bars show the variability of $m_1(m_0)$ and $m_2(m_0)$ with different spinodal-like patterns (not shown). Solid lines correspond to the expected values for a Gaussian random field (GRF)~\cite{Mantz_2008}.}
    \label{fig:figure2}
\end{figure}

Patterns qualitatively similar to the experimental ones are reproduced by a minimal phase-field (PF) model, which accounts for the energetics of surfaces, favouring surface smoothing by surface diffusion and the competing relaxation of biaxial (misfit) strain, favouring surface corrugations as in the context of the ATG instability~\cite{AsaroMT1972,GrinfeldJNS1993,SrolovitzAMETAL1989}. 
The considered PF model describes these contributions and tracks the surface $\Sigma$ of the film as the 0.5 level set of a continuous variable $\varphi$ set to $1$ in Si$_{1-x}$Ge$_{x}$ and UT-SOI, and $0$ in vacuum, while changing smoothly in between over a length $\epsilon$~\cite{Ratz2006,Li2009}. 
The model reads
\begin{equation}
\begin{split}
\partial_t \varphi &=D\nabla\cdot\left(\frac{1}{\epsilon}M(\varphi)\nabla\omega\right)\\
g(\varphi) \omega&=\gamma \left( -\epsilon \nabla^2 \varphi + \frac{1}{\epsilon} F^\prime(\varphi)\right) + \frac{\partial \rho_{e}(\varphi,\boldsymbol{\varepsilon})}{\partial \varphi}
\end{split}
\label{eq:pf}
\end{equation}
with $\omega$ the chemical potential, $M(\varphi)=36\varphi^2(1-\varphi)^2$ a mobility function, $F(\varphi) = 18\varphi^2(1-\varphi)^2$ a double-well potential, $g(\varphi) = 30 \varphi^2 (1- \varphi)^2$ a stabilisation function, $D$ a diffusion coefficient (incorporated in the timescale $t'=t/D$)~\cite{Ratz2006,Li2009}, and $\gamma$ the isotropic surface-energy density. $\partial \rho_{e}(\varphi,\boldsymbol{\varepsilon}) / \partial \varphi$ encodes the elastic-energy contribution (towards roughening) to the chemical potential, with $\rho_{e}(\varphi,\boldsymbol{\varepsilon})=\boldsymbol{\varepsilon}:\boldsymbol{\sigma}(\varphi,\boldsymbol{\varepsilon})$ the elastic energy density, $\boldsymbol{\sigma}(\varphi,\boldsymbol{\varepsilon})=\mathbf{C}(\varphi):\boldsymbol{\varepsilon}$ the stress field, $\boldsymbol{\varepsilon}=(1/2)(\nabla \mathbf{u} + (\nabla \mathbf{u})^{\rm T})+h(\varphi)\varepsilon_{\rm 0}\mathbf{I}$ the strain field, $\mathbf{u}$ the displacement w.r.t the relaxed state, $\mathbf{C}(\varphi)$ the elastic constant tensor and $h(\varphi)$ an auxiliary function vanishing in the vacuum phase. $\varepsilon_{\rm 0}$ is the effective misfit strain between Si$_{1-x}$Ge$_{x}$ and UT-SOI (residual in the presence of plastic relaxation). $\boldsymbol{\varepsilon}$ is determined by solving the mechanical equilibrium equation $\nabla \cdot \boldsymbol{\sigma}(\varphi,\mathbf{u})=0$ \cite{Ratz2006,Bergamaschini2016}. Due to its very small relative thickness, the Si layer is not explicitly considered, but its presence is encoded in $\varepsilon_{\rm 0}$ by strain accumulation. For the asymptotic analysis $\epsilon \to 0$ we refer to \cite{Ratz2006,Voigt2016,SalvalaglioDDCH}. The amorphous buried oxide is modeled through a no-flux boundary condition for $\varphi$~\cite{Jiang2012,Naffouti2017} allowing for dewetting and enforcing a $90^\circ$ contact angle, without loss of generality~\cite{Naffouti2017,Backofen2019}. Other effects, such as plastic relaxation and intermixing, can be treated as changes in the eigenstrain in a mean-field approximation~\cite{SalvalaglioAPL2018}. We express the system size (such as thickness $h$, and lateral extension $L$) in terms of the characteristic ATG length $\ell=\gamma/\rho_{\rm flat}$, with $\rho_{\rm flat}$ the elastic energy density of a flat film under biaxial strain~\cite{AsaroMT1972,GrinfeldJNS1993,SrolovitzAMETAL1989}. 
Numerical simulations are performed using the parallel finite element toolbox AMDiS~\cite{Vey2007,WitkowskiACM2015} with adaptive time steps and mesh refinement (see also \cite{Sal2015b,Bergamaschini2016,Backofen2019} and S.M.). 

In agreement with the ATG instability, well describing the early evolution of a perturbed film, the corrugation has a characteristic wavelength $\ell$ and amplifies over time (Fig.~\ref{fig:figure2}) \cite{FOOTNOTE}. When the trenches reach the BOX, dewetting sets in everywhere and almost simultaneously in the film. Despite the simplified model, the main features of the dewetting dynamics are qualitatively reproduced for different film thicknesses, annealing time and temperature (accounted for by the diffusion coefficient $D$ depending on temperature by an Arrhenius law~\cite{Mullins1957}).

The morphology and nature of disorder of the dewetted structures in experiments and simulations are analyzed by Minkowski functionals, that quantify the topological properties and, in turn, spatial features of 2D patterns
\cite{MIN1989,Becker2003,SCH2014}.  We consider averaged Minkowski functionals~\cite{Mantz_2008} (see also S.M.) $m_i(\bar{\rho})=(1/|\Omega|)M_i(\mathcal{B}_{\bar{\rho}})$ 
of thresholded 8-bit gray-scale images representing the thickness of samples by a space-dependent field $\rho(\mathbf{p}_{ij})\in[0,255]$, with $\mathbf{p}_{ij}$ the coordinate of image pixels, over a region with extension $|\Omega|$. $\bar{\rho}$ is a given threshold defining a binary image as $\mathcal{B}_{\bar{\rho}}=\Theta(\rho(\mathbf{p}_{ij})-\bar{\rho})$ and $\Theta$ the Heavyside function.
$m_0(\bar{\rho})$ corresponds to the fraction of $|\Omega|$ occupied by the non-zero region in $\mathcal{B}_{\bar{\rho}}$, as $M_0(\mathcal{B}_{\bar{\rho}})$ gives the area occupied by $\mathcal{B}_{\bar{\rho}}=1$. $m_1(\rho)$ represents an average of the boundary length $U$ between the regions $\mathcal{B}_{\bar{\rho}}=1$ and $\mathcal{B}_{\bar{\rho}}=0$, as $U=2 \pi M_1(\mathcal{B}_{\bar{\rho}})$. Similarly, $m_2(\rho)$ corresponds to the averaged Euler characteristic $\chi$, as $\chi=\pi M_2(\mathcal{B}_{\bar{\rho}})$. The plot of $m_{1,2}(m_0)$ with the encoded averaging, provides results independent of image saturation and contrast~\cite{Mantz_2008}. 

$m_{1,2}(m_{0})$ for representative experimental and simulated patterns almost overlap for most of the $m_0$ range (Fig.~\ref{fig:figure2}). Different experimental connected and disconnected structures follow a similar behavior (see the range of variation highlighted by dashed error bars in Fig.~\ref{fig:figure2}(c)). $m_{1,2}(m_{0})$ assesses the deviation from a Gaussian random field (solid lines)~\cite{Mantz_2008}. This deviation points to non-linearity of the underlying dynamics and correlations in the resulting patterns (for a similar analysis see~\cite{Galinski2014}).

\begin{figure}[t]
    \centering
    \includegraphics[width=\linewidth]{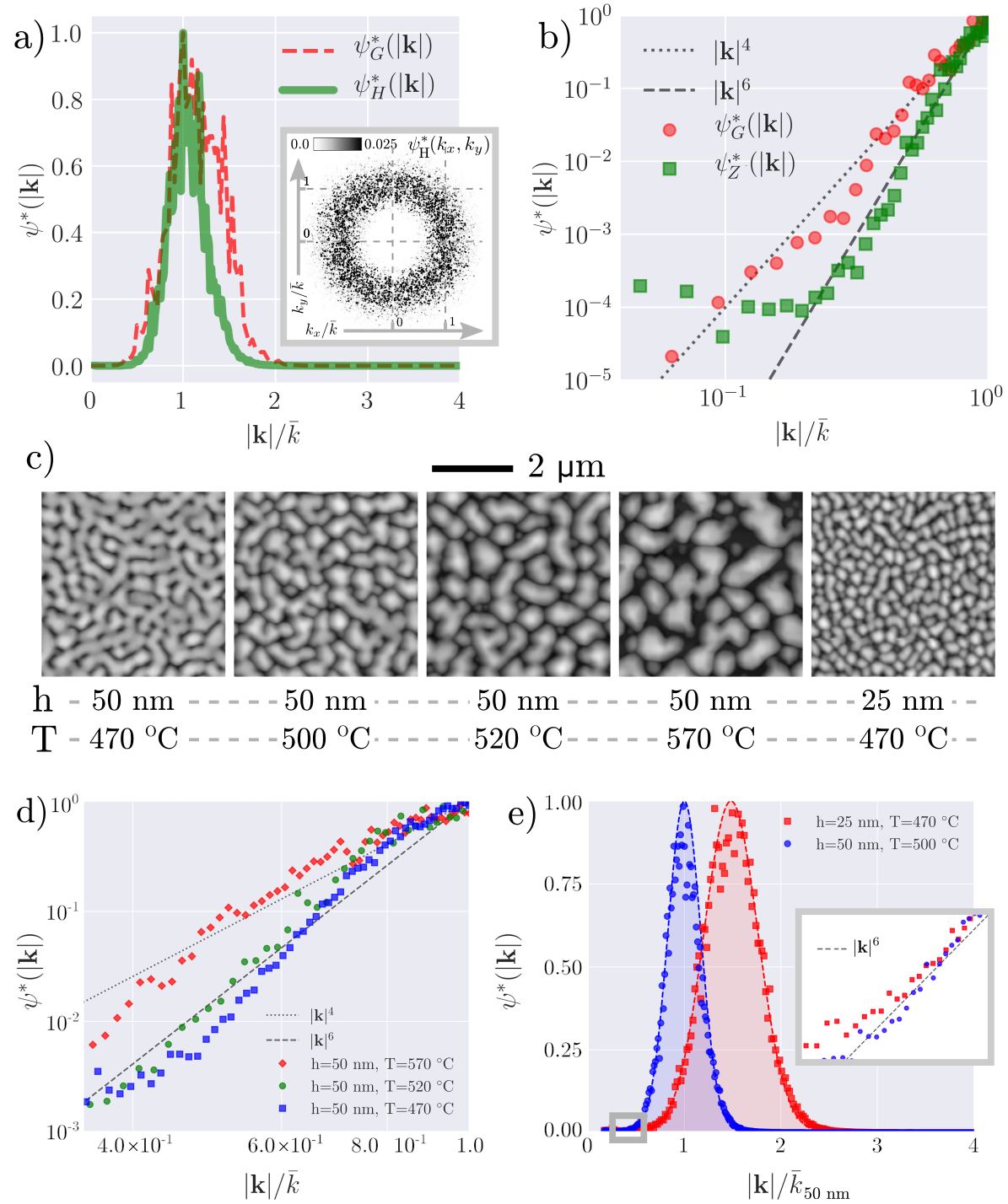}   
    \caption{\textit{Hyperuniformity and tuning of the disorder}: (a) Radial spectral density $\psi^*(|\mathbf{k}|)$  for the pattern of Fig.~\ref{fig:figure2}(c) ($\psi_{H}^*(|\mathbf{k}|)$) and a numerically generated GRF ($\psi_{G}^*(|\mathbf{k}|)$), with the inset showing $\psi_{H}^{*}(k_x,k_y)$. (b) Logarithmic plot of data in panel (a) highlighting the decay of $\psi^*(|\mathbf{k}|)$ for $|\mathbf{k}|\rightarrow 0$. Dashed and dotted lines are the trend of $|\mathbf{k}|^{4}$ and $|\mathbf{k}|^{6}$ respectively.
    (c) SEM images of Ge ($h$=50 nm and 25 nm) deposited on UT-SOI and annealed at different temperature. (d) Logarithmic plot of experimental $\psi_{Z}^*(|\mathbf{k}|)$ for similar samples annealed at different temperature (see panel (c), corresponding large view 40~$\mu$m $\times$ 40~$\mu$m AFM images have been used). (e) Comparison of $\psi_{Z}^*(|\mathbf{k}|)$ of the patterns in the two last panels of (c).}
    \label{fig:figure3}
\end{figure}

Based on the theoretical prediction in references \onlinecite{Torquato2016,Ma2017} we further assess the correlation properties of the dewetted structures by looking at the spectral density $\psi^*(|\mathbf{k}|)$ with $\mathbf{k}=(k_x,k_y)$ (Fig.~\ref{fig:figure3}(a)-(b)) of their correlation function $\psi(x,y)$ obtained from a given height profile $Z(x,y)$ (inset of Fig.~\ref{fig:figure2}(c)) and, in particular, at its decay for small wavenumbers (long wavelength)~\cite{Torquato2016,Ma2017,Gruhn_2016}. Patterns can be considered to be hyperuniform if $\psi^*(|\mathbf{k}|/\bar{k}) = (|\mathbf{k}|/\bar{k})^{\beta}$  and $\beta \geq 4$ with $|\mathbf{k}|/\bar{k} \rightarrow 0$ ($\bar{k}$ is the position of the maximum in the reciprocal space). $\beta = 4$ corresponds to a Gaussian random field whereas larger values occur in the presence of correlations leading to stronger suppression of long wavelengths oscillations~\cite{KLA2019,TOR2018}. For the morphology shown in the inset of Fig.~\ref{fig:figure2} (c) we obtain $5<\beta<6$ for $0.1 < |\mathbf{k}|/\bar{k} < 1$ (Fig.~\ref{fig:figure3} (b)). In most of the analyzed patterns $\beta \sim 6$ when considering sufficiently large images 
Whereas hyperuniform structures present a ratio $H=\psi^*(0^+)/\psi^*(\bar{k})$ strictly equal to zero, a $H\le 10^{-4}$ (as in Fig.~\ref{fig:figure3} (c)) is accepted to be a fingerprint of hyperuniformity, at least for real systems~\cite{KLA2019,TOR2018} and in the presence of other arguments supporting long-range interactions and correlations~\cite{Jiao2014}. Indeed, inherent spurious effects arising from image thresholding~\cite{Ma2017} affect the steep decrease of $\psi^*(|\mathbf{k}|)$ for smaller $\mathbf{k}$ (Fig.~\ref{fig:figure2}(d) and (e)). We thus consider the emergent patterns by spinodal solid-state dewetting as hyperuniform. 

The degree of hyperuniformity can be tuned by considering different temperature and thus, different stages during the dewetting process (see Fig.~\ref{fig:figure3}(d), obtained by considering samples with $x$ = 1, thickness of 50~nm and changing the annealing condition). Larger and disconnected structures behave more like a GRF ($\beta \sim 4$), whereas for more connected structures a marked hyperunifomity emerges ($\beta \sim 5-6$). We can also replicate morphologies featuring a similar hyperuniform character (e.g., $\beta \sim 6$, Fig.~\ref{fig:figure3}(e)) but different $\bar{k}$, i.e. peak position of $\psi_{Z}^*(|\mathbf{k}|)$ obtained by dewetting Ge layers having different initial thickness (Fig.~\ref{fig:figure3} (c), respectively 50~nm, T = 500$\celsius$ and 25~nm, T = 470$\celsius$). This demonstrates a size tuning of structures exhibiting similar correlations.

In conclusion, we identified an elasticity-driven instability in strained, thin films undergoing solid-state dewetting, enabling the formation of spinodal, monocrystalline patterns that exhibit effective hyperuniformity. The  process is scalable, as it does not depend on the sample size and, more importantly, adjustable: setting layer thickness and annealing cycle provides structures with different topology (connected to disconnected), composition (from Si$_{0.7}$Ge$_{0.3}$ to pure Ge), and size (from hundreds of nm to tens of $\mu$m). Examples are provided at a glance in the S.M.. The class of hyperuniform materials, including so far various natural and artificial systems, such as, polymer-grafted nanoparticles~\cite{CHR2017}, polymer melt~\cite{XU2016}, colloid deposition~\cite{PIE2018}, maximally random jammed packing~\cite{ATK2016}, clusters of nanoparticles~\cite{de2015toward} and foams~\cite{ricouvier2019foam} has then been extended to technology-relevant materials. Furthermore, a remarkable advantage of spinodal solid-state dewetting for hyperuniform materials over conventional processes at critical points~\cite{Ma2017,ma2020} is the possibility to freeze the system in a desired condition, thus setting at will its properties. 

The choice of a UT-SOI substrate and Si$_{1-x}$Ge$_{x}$ alloys, which are common and cost-effective materials, opens the path to fruitful industrialization as a realistic platform for electricity, light or matter management. Moreover, the fabrication of hyperuniform devices has been attempted with more or less cumbersome top-down approaches over limited extensions~\cite{CAS2017,MUL2014,FRO2016,FRO2017} and, in this respect, the reported dewetting process is more convenient, as it can be implemented in a single fabrication step and results in atomically smooth mono-crystals having, for instance, superior electronic figures of merits with respect to rough, etched materials~\cite{BOL19}. 

\section*{Acknowledgements}
We acknowledge fruitful discussions with Marian Florescu (University of Surrey), Riccardo Sapienza (Imperial College London) and Dominique Chatain (CINAM, Marseille). This research was funded by the EU H2020 FET-OPEN project NARCISO (ID: 828890). J. W. and J-B.  C. acknowledge the European Research Council (ERC, grant agreement No 723241). A. B. and M. A. acknowledge PRCI network ULYSSES (ANR-15-CE24-0027-01) funded by the French ANR agency. We gratefully acknowledge the computing time granted by Julich Supercomputing Centre (JSC) within the Project no. HDR06, and by ZIH at TU Dresden. We acknowledge the NanoTecMat platform of the IM2NP institute of Marseilles
and the microscopy center of Aix-Marseille University CP2M.

%

\pagebreak
\onecolumngrid
\newpage
\begin{center}
  \textbf{\large \hspace{5pt} SUPPLEMENTAL MATERIAL \\ \vspace{0.2cm} Hyperuniform monocrystalline structures by spinodal solid-state dewetting}\\[.2cm]
  Marco Salvalaglio,$^{1,2,*}$ Mohammed Bouabdellaoui,$^{3}$, Monica Bollani$^{4,\dagger}$, Abdennacer Benali$^3$, Luc Favre,$^3$ \\ Jean-Benoit Claude$^5$, Jerome Wenger$^5$, Pietro de Anna$^6$, Francesca Intonti$^7$, Axel Voigt$^{1,8}$, Marco Abbarchi$^{3,\ddagger}$, \\[.1cm]
  {\itshape \small
  ${}^1$Institute  of Scientific Computing,  TU  Dresden,  01062  Dresden,  Germany,
  \\ 
  ${}^2$Dresden Center for Computational Materials Science (DCMS), TU  Dresden,  01062  Dresden,  Germany
 \\
  ${}^3$Aix Marseille Univ, Universit\'e de Toulon, CNRS, IM2NP, Marseille, France,
 \\
  ${}^4$Istituto di Fotonica e Nanotecnologie-Consiglio Nazionale delle Ricerche,\\ Laboratory for Nanostructure Epitaxy and Spintronics on Silicon, Via Anzani 42, 22100 Como, Italy,
  \\
  ${}^5$Aix Marseille Univ, CNRS, Centrale Marseille, Institut Fresnel, 13013 Marseille, France,
  \\
  ${}^6$Institut des Sciences de la Terre, University of Lausanne, Lausanne 1015, Switzerland,
  \\
  ${}^7$LENS, University of Florence, Sesto Fiorentino, 50019, Italy,
  \\}
\vspace{0.5cm}
\end{center}
\twocolumngrid

\setcounter{equation}{0}
\setcounter{figure}{0}
\setcounter{table}{0}
\setcounter{page}{1}
\renewcommand{\theequation}{S\arabic{equation}}
\renewcommand{\thefigure}{S\arabic{figure}}
\renewcommand{\bibnumfmt}[1]{[S#1]}
\renewcommand{\citenumfont}[1]{S#1}

\section{Methods}

\begin{figure*}[t]
    \centering
   \includegraphics[width=\textwidth]{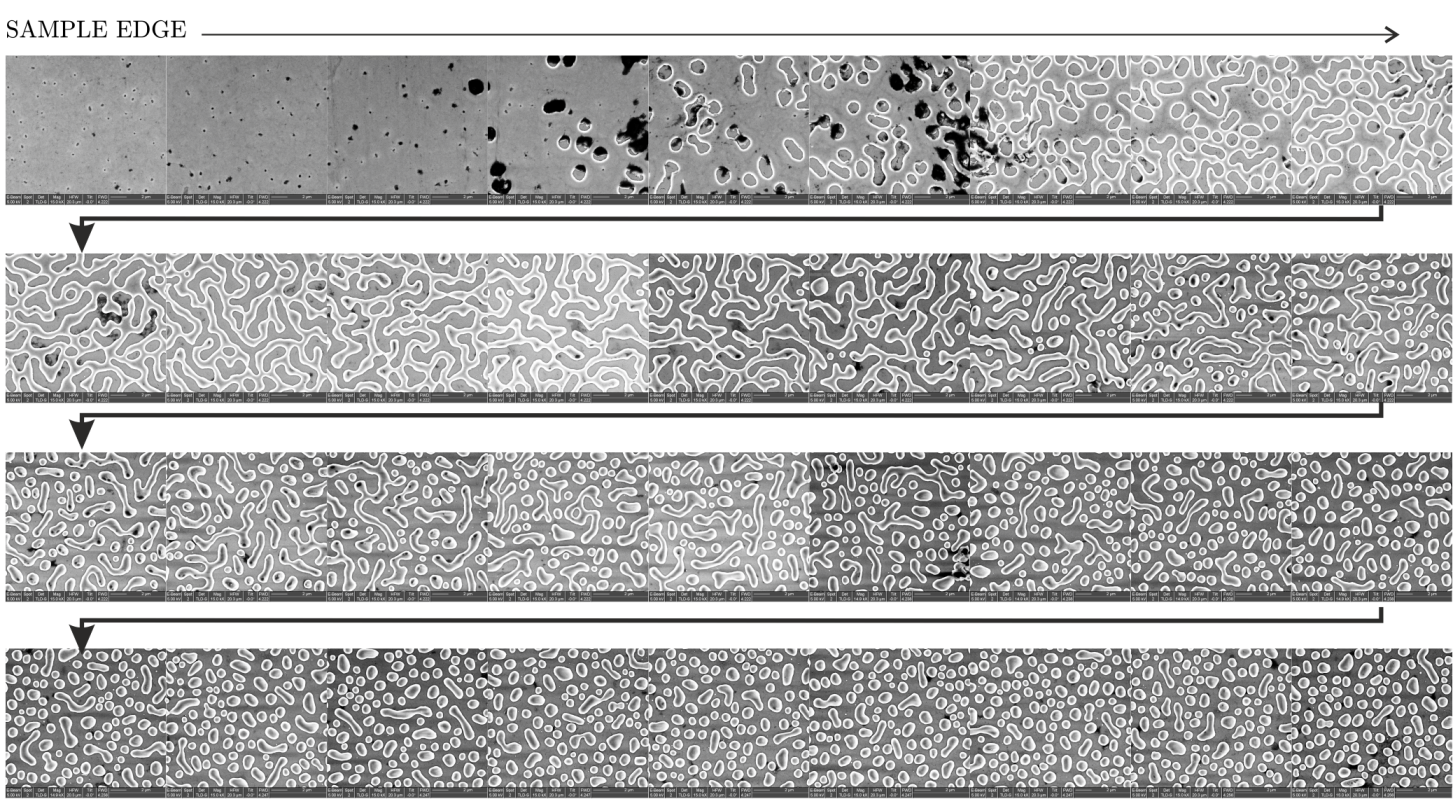}   
    \caption{Scanning electron micrographs (SEM) displaying the morphological evolution (from the edge of the sample, left panel, towards its center, right panel) of 150~nm Si$_{0.7}$Ge$_{0.3}$ on UT-SOI after 4 hours annealing at 800$\celsius$. Another example from a different sample is provided as a separate high-resolution Supplementary File 1 where optical microscopy shows that the change in morphology occurs over about 0.5~mm from the edge.}
    \label{fig:SIfigure1}
\end{figure*}

 All dewetted structures are obtained starting from the same ultra-thin, silicon-on-insulator wafer (UT-SOI, double-polished, oriented along the (001) direction): 14 nm of Si on 25 nm of SiO$_{2}$ (buried oxide, BOX) on bulk Si. The UT-SOI is first cleaned in a HF solution ($\sim$5$\%$) for 20'' in a glove box under nitrogen atmosphere to remove the native oxide. Then, it is introduced in the ultra-high vacuum of a molecular beam epitaxy reactor (MBE) (static vacuum $\sim$10$^{-10}$~Torr). The samples undergo a first flash annealing at 600~$^{\circ}$C for 30' to remove any trace of native oxide. SiGe alloys of variable thickness are deposited on the UT-SOI and annealed at temperatures ranging from 400~$^{\circ}$C to 800~$^{\circ}$C to tune the dewetted morphologies (e.g., island or connected nano-architectures). After solid-state dewetting, the samples are investigated via atomic force microscopy (AFM) in non-contact mode, high resolution scanning electron microscopy (SEM), and transmission electron microscopy (TEM).

\section{Tuning of morphologies and sizes}

The dewetted islands' morphology can be tuned in a wide range of SiGe thickness and Ge content, adjusting annealing temperature and time. Figure 1 in the main text considers the case of temperature change for a given deposited thickness and annealing time. It exploits a temperature gradient from the sample edge towards its center present in the molecular beam reactor. In Fig.~\ref{fig:SIfigure1}, a more precise characterization of the same sample is provided. This is obtained from an individual sample observing the evolution of the layer morphology from the sample edge towards its center. In this case, the nominal annealing temperature was 800$^{\circ}$C, as measured with a pyrometer at the center of the sample. Owing to the holder that clamps the sample at its edges, the temperature on the sides is lower than at the sample center. This allows us to monitor the effect of temperature while not changing all the other conditions (layer thickness and annealing time), and it has been exploited to show morphologies after dewetting at different temperatures in many experiments involving conventional solid-state dewetting processes \cite{SINaffouti2017,SIBOL19}. Provided that the dewetting at the sample edges is barely visible (only a few holes form), we estimate the temperature at the edge of about 750$^{\circ}$C compared with the morphologies obtained at the center of the sample with this temperature as measured by the pyrometer. We deduce this value by comparing this morphology with that one found at the center of other samples grown at lower temperature.   

From optical microscopy on a similar sample (200~nm Si$_{0.7}$Ge$_{0.3}$ on UT-SOI after 4 hours annealing at 800$\celsius$), by scanning over 2~mm  from the edge, we estimate that the morphology change occurs within 1-2~mm. In the central part of the sample the morphology of the nano-architectures is the same, providing homogeneous properties. This image is provided as a separate high-resolution Supplementary File 1, allowing for zooming-in to see better the fine details of the dewetted structures. It was obtained with a Leica optical microscope with a 100$\times$ magnification objective lens in dark-field configuration.   

\begin{figure}[b!]
    \centering
   \includegraphics[width=1\columnwidth]{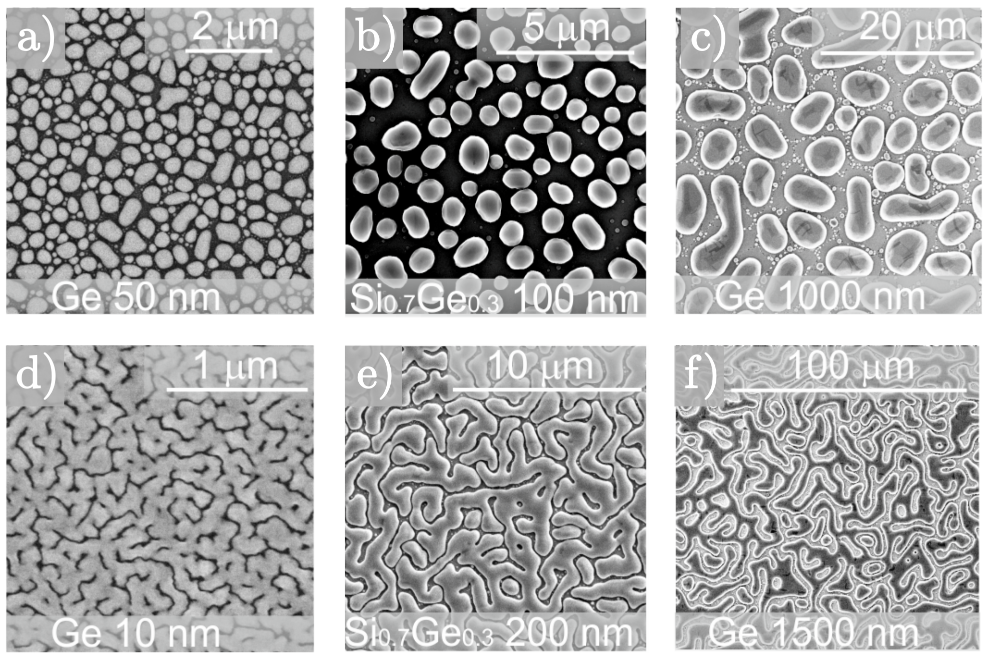} 
   \caption{\textit{Size and morphology tuning}. SEM images of Si$_{1-x}$Ge$_{x}$ layers of thickness $h$ deposited on UT-SOI and annealed for a time $t$ at temperature $T$: a) $x$=1, $h$=50~nm, $t$=45~min, $T$=750$\celsius$; b) $x$=0.3, $h$=100~nm, $t$=150~min, $T$=800$\celsius$; c) $x$=1, $h$=1000~nm, $t$=4~h, $T$=800$\celsius$;  d)  $x$=1, $h$=10~nm, $t$=30~min, $T$=450$\celsius$; e)  $x$=0.3, $h$=200~nm, $t$=1~h, $T$=800$\celsius$. f)  $x$=1, $h$=1500~nm, $t$=4~h., $T$=800$\celsius$.   
}
    \label{fig:figure4}
\end{figure}

The dewetting process and, in turn, the resulting bottom-up fabrication technique is scalable as it does not depend on the sample size. It is also adjustable: by an appropriate choice of annealing temperature and time we can obtain structures in a wide range of sizes (e.g., depositing Si$_{1-x}$Ge$_{x}$ thicknesses from 5 to 2000~nm via MBE), Ge-content (from x = 0.3 up to 1) and morphology (e.g., holes, islands or connected structures).  At a glance, a few examples are provided in Fig.~\ref{fig:figure4} showing typical size ranging from hundreds of nanometers to several microns for islands (Fig.~\ref{fig:figure4} (a)-(c)), as well as for connected structures (Fig.~\ref{fig:figure4} (d)-(f)). 

\section{Dislocations in large structures from thick  $Si_{0.7}Ge_{0.3}$ dewetting}

\begin{figure*}[!ht]
    \centering
   \includegraphics[width=\textwidth]{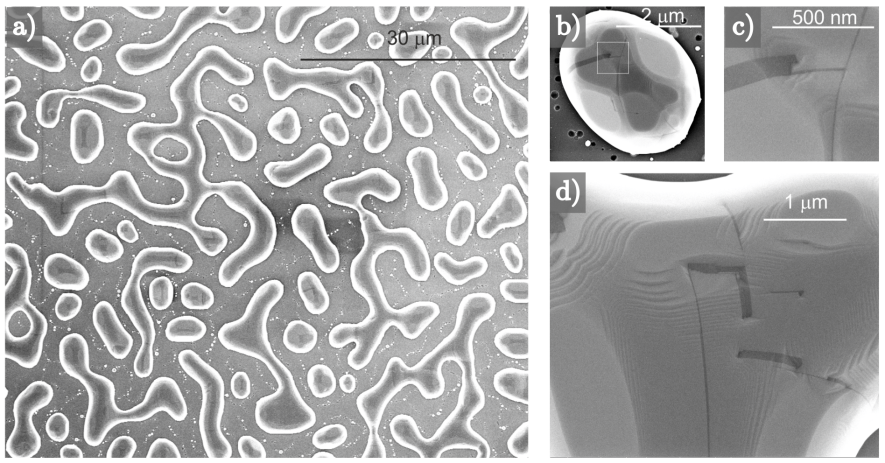}   
    \caption{a) Scanning electron micrograph of 500~nm Ge on UT-SOI after 6 hours annealing at 800$^{\circ}$C. b) Detail of a dewetted island with large facets. c) Zoom of the area highlighted by a white square in b). d) Detail of facets, terraces and step bunching.\vspace{0.5cm}}
    \label{fig:SIfigure2}
\end{figure*}

The presence of strain between thick  Si$_{0.7}$Ge$_{0.3}$ layers and the UT-SOI potentially leads to 
plastic relaxation and formation of dislocation.
This is also evident from SEM images from large islands, see Figure~\ref{fig:SIfigure2} a), which shows the case of 500~nm Ge on UT-SOI after 6 hours annealing at 800$^{\circ}$C. The overall morphology is very similar to that observed in thinner samples (as shown in the main text). The large size of the islands allows us to highlight the presence of facets, terraces, step bunching, and the presence of a dislocation emerging at the surface of the island (Figure~\ref{fig:SIfigure2} b)-d).

\section{Atomic force microscopy of dewetted samples}

\begin{figure*}[ht!]
    \centering
   \includegraphics[width=\textwidth]{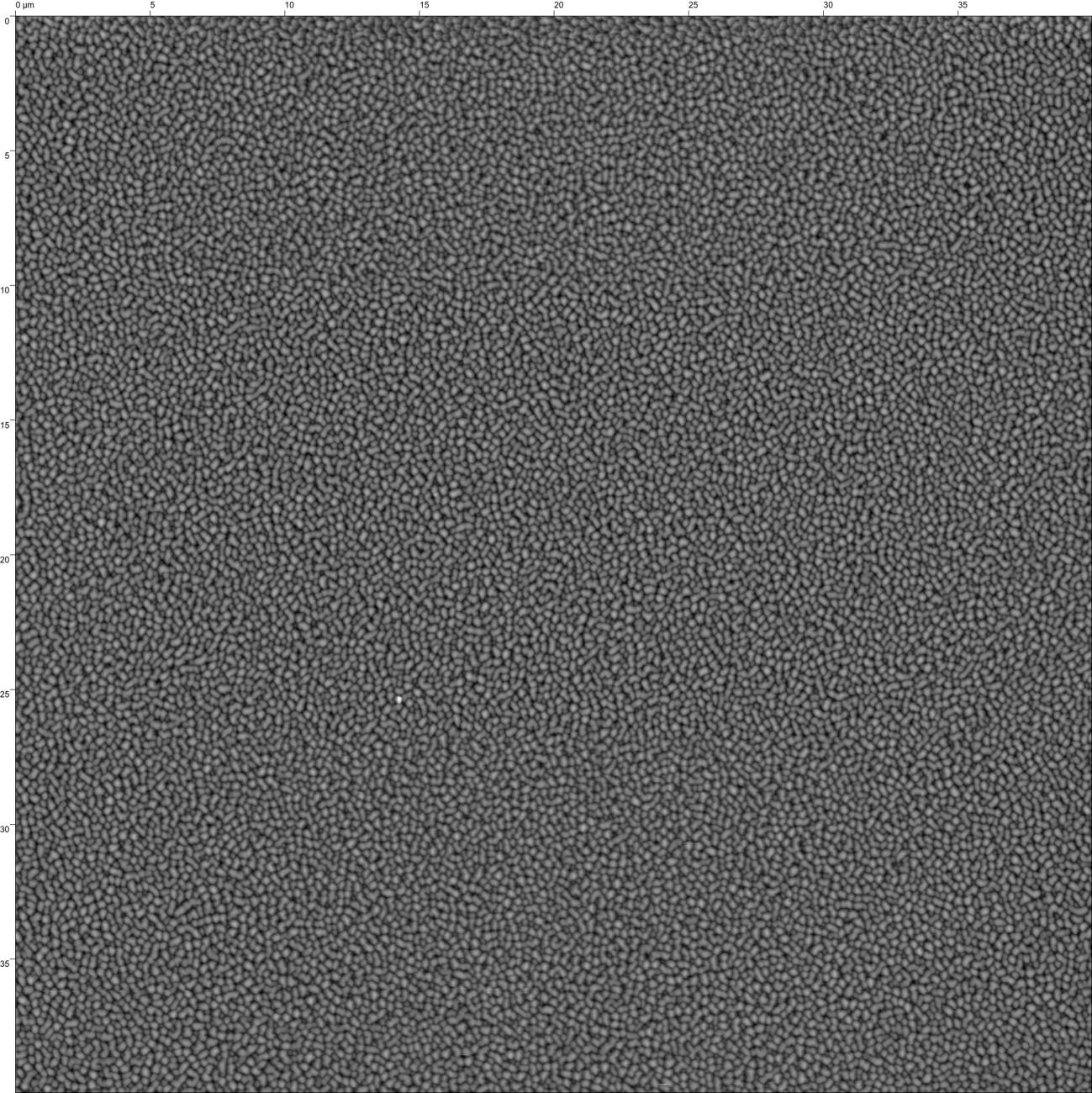}   
    \caption{Large-scale atomic force microscope image, 40$\times$40~$\mu$m$^{2}$ with a 2048$\times$2048 points, of  25~nm Ge deposited on UT-SOI and annealed at 480$^{\circ}$C for 45 minutes. This image was used for Minkowski analysis of experimental patterns and for the hyperuniformity analysis.}
    \label{fig:SIfigure3}
\end{figure*}

For a precise assessment of the disorder properties (e.g., hyperuniform character) of the dewetted structures, we use AFM, high-resolution and large images. The characterization is performed over 40$\times$40~$\mu$m$^{2}$ with 2048$\times$2048 points. A 25~nm Ge deposited on UT-SOI, annealed at 480$^{\circ}$C for 45 minutes, is shown in Figure~\ref{fig:SIfigure3}. AFM images of same size and resolution were performed on each sample shown in Fig.3 of the main text in order to estimate correlation function and corresponding spectral density. 

\section{Phase-Field model}

A phase-field model that accounts for both surface and elastic energy is considered. It is based on the model introduced in Refs.~\cite{SIRatz2006}, where a continuous order parameter $\varphi\equiv\varphi(\mathbf{x})$ that is set to be $\varphi(\mathbf{x})=1$ for $\mathbf{x}$ in the solid phase (Si$_{1-x}$Ge$_{x}$, UT-SOI), 
$\varphi(\mathbf{x})=0$ for $\mathbf{x}$ in the vacuum, and with a continuous transition in between well described by
\begin{equation}
\varphi(\mathbf{x})=\frac{1}{2}\left[1-\tanh \left(\frac{3 d(\mathbf{x})}{\epsilon}\right)\right].
\end{equation}
$\epsilon$ is the interface thickness between solid and vacuum phase, $d(\mathbf{x})$ corresponds to the signed distance from the interface $\Sigma$, corresponding then to 0.5 level set of $\varphi$. The model is based on the free-energy functional
\begin{equation}
F[\varphi]=\int_\Omega \gamma \left( \frac{\epsilon}{2} |\nabla \varphi|^2  + \frac{1}{\epsilon}F(\varphi) \right) d\mathbf{x}
+\int_\Omega \rho_{\rm e}(\varphi,\boldsymbol{\varepsilon}) d\mathbf{x},
\label{eq:energy}
\end{equation}
with $\Omega$ the 3D simulation domain and $\mathbf{x} \in \Omega$. The first integral in Eq.~\eqref{eq:energy} accounts for the surface energy with $F(\varphi)=18\varphi^2(1-\varphi)^2$ and $\gamma$ the isotropic surface energy. The second integral in Eq.~\eqref{eq:energy} accounts for the elastic energy with 
\begin{equation}
\rho_{\rm e}(\varphi,\boldsymbol{\varepsilon}) = \upmu h(\varphi) \sum_{ij}\varepsilon_{ij}^2+\frac{\uplambda }{2}h(\varphi)\left(\sum_{i}\varepsilon_{ii} \right)^2
\label{eq:elasdensity}
\end{equation}
the elastic-energy density for an isotropic media, $\uplambda$ and $\upmu$ the Lame constants, $h(\varphi)=\varphi^3(6\varphi^2-15\varphi+10)$ a function smoothly vanishing when approaching the vacuum phase. $\boldsymbol{\varepsilon}$ is the strain tensor
\begin{equation}
\boldsymbol{\varepsilon}=\frac{1}{2}\left[\nabla \mathbf{u} + (\nabla \mathbf{u})^{\rm T}\right]-\boldsymbol{\varepsilon}_0,
\label{eq:strain}
\end{equation}
with displacement $\mathbf{u}$ with respect to the relaxed state and a term $\boldsymbol{\varepsilon}_0$, which encodes the permanent deformation due to the mismatch between the lattice constant of the epilayer (Si$_{1-x}$Ge$_{x}$), $a_{\rm e}$, and of the substrate (UT-SOI) $a_{\rm s}$ such as $\varepsilon_{\rm 0}=(a_{\rm s}-a_{\rm e})/a_{\rm s}$, and 
\begin{equation}
\boldsymbol{\varepsilon}_0=-h(\varphi)\varepsilon_{\rm 0}\mathbf{I}.
\end{equation}  
The strain tensor entering Eq.~\eqref{eq:elasdensity} is determined by assuming mechanical equilibrium, namely by solving
\begin{equation}
\nabla \cdot \boldsymbol{\sigma}=0,
\label{eq:mecheq}
\end{equation}
with 
\begin{equation}
\boldsymbol{\sigma}=\uplambda \text{Tr}(\boldsymbol{\varepsilon})\mathbf{I}+2\upmu \boldsymbol{\varepsilon}.
\end{equation}
The evolution law for $\varphi$ is given by the conservative gradient flow
\begin{equation}
\frac{\partial \varphi}{\partial t}=D\nabla \cdot \left( \frac{1}{\epsilon} M(\varphi) \nabla \omega \right),
\label{eq:dphidt}
\end{equation}
with $M(\varphi)=36\varphi^2(1-\varphi)^2$ a degenerate mobility and $D$ the effective diffusion coefficient. The latter expected to depend on temperature by an Arrhenius law, encoding the activation of surface diffusion at high temperature. Without loss of generality it is here set to 1. The effect of $D$ is incorporated in the timescale such as $ t'\rightarrow t/D$.
$\omega={\delta F}/{\delta \varphi}=\omega_{\rm s}+\omega_{\rm e}$ is determined by
\begin{equation}
    \begin{split}
        g(\varphi)\omega_{\rm s} = & \gamma \left( -\epsilon \nabla^2 \varphi + \frac{1}{\epsilon} F'(\varphi) \right)
        \\
        g(\varphi)\omega_{\rm e} = & \frac{\partial\rho_{e} (\varphi,{\varepsilon})}{\partial \varphi}
\end{split}
\end{equation}
with $g(\varphi)=30\varphi^2(1-\varphi)^2$ an additional degeneracy improving the approximation of surface diffusion for relatively large $\epsilon$ \cite{SIRatz2006,SIVoigt2016,SISalvalaglioDDCH}. 
A contact angle of $90^\circ$ at the substrate (BOX) is enforced by a no-flux boundary condition for the order parameter $\varphi$ \cite{SIBackofen2019}. 

The model presented so far allows for reproducing the main experimental features reported in this work. Model extensions to consider the substrate or a second solid phase \cite{SIWise2005,SIAlbani2016,SISalvalaglioAPL2018}, alloys \cite{SIBackofen2014}, surface-energy anisotropy leading to surface faceting \cite{SITor2009,SISal2015b,SISalvalaglioNRL2017} and different contact angles \cite{SIBackofen2019} can be considered. Simulations were performed by employing the parallel Finite Element toolbox AMDiS \cite{SIVey2007,SIWitkowskiACM2015}, allowing for time adaptivity and mesh refinement at the solid-vacuum interface. The equations governing the evolution of $\varphi$ are solved as a system of two second-order partial differential equation, namely for $\varphi$ and $\omega$, and discretized by a semi-implicit integration scheme (see Ref.~\cite{SIRatz2006,SISal2015b,SIBackofen2019}). Periodic boundary conditions are imposed at the lateral boundaries for $\varphi$, while no-flux Neumann boundary conditions are imposed elsewhere. Strain fields at equilibrium are computed by solving Eq.~\eqref{eq:strain} and \eqref{eq:mecheq} for $\mathbf{u}$ at each timestep. Periodic boundary conditions are imposed along in-plane directions, while a Dirichlet boundary condition with $\mathbf{u}=0$ is imposed at the bottom of the simulation domain.

\begin{figure}
    \centering
    \includegraphics[width=\linewidth]{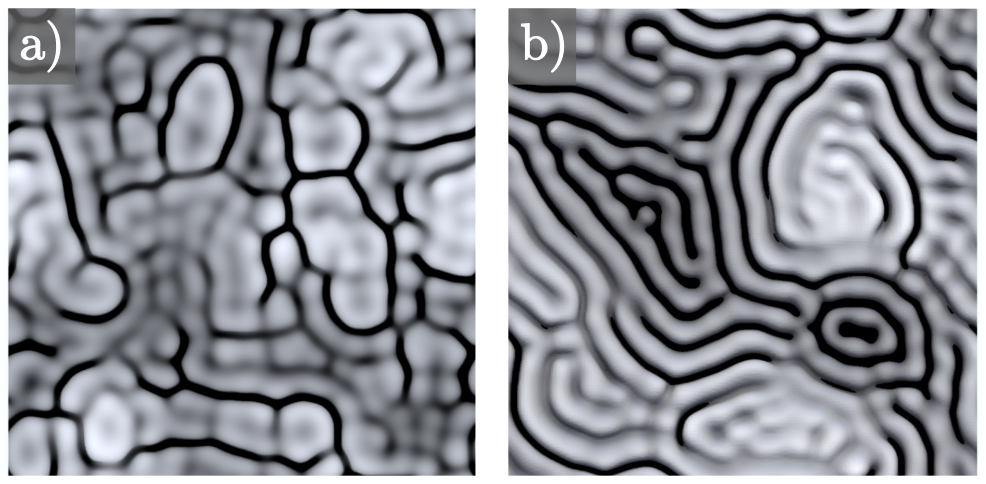}   
    \caption{Different simulated pattern depending on the wavelengths encoded in the initial perturbation, $\lambda_i$ with respect to the characteristic length of the ATG instability $\ell$. a) $\lambda_i < \ell$, b) $\lambda_i > \ell$.}
    \label{fig:SIfigure4}
\end{figure}

Fig.~\ref{fig:SIfigure4} shows different outcomes according to the wavelengths encoded in the initial perturbation. In agreement with the literature \cite{SIBergamaschini2016}, wavelengths smaller than the characteristic length of the ATG instability must be considered for the proper development of the instability for an unpatterned film. The outcome of simulations with initial perturbation containing large wavelengths only may suggest a way to obtain striped patterns by proper preparation of the substrate.

\begin{figure}[!b]
    \centering
    \includegraphics[width=\linewidth]{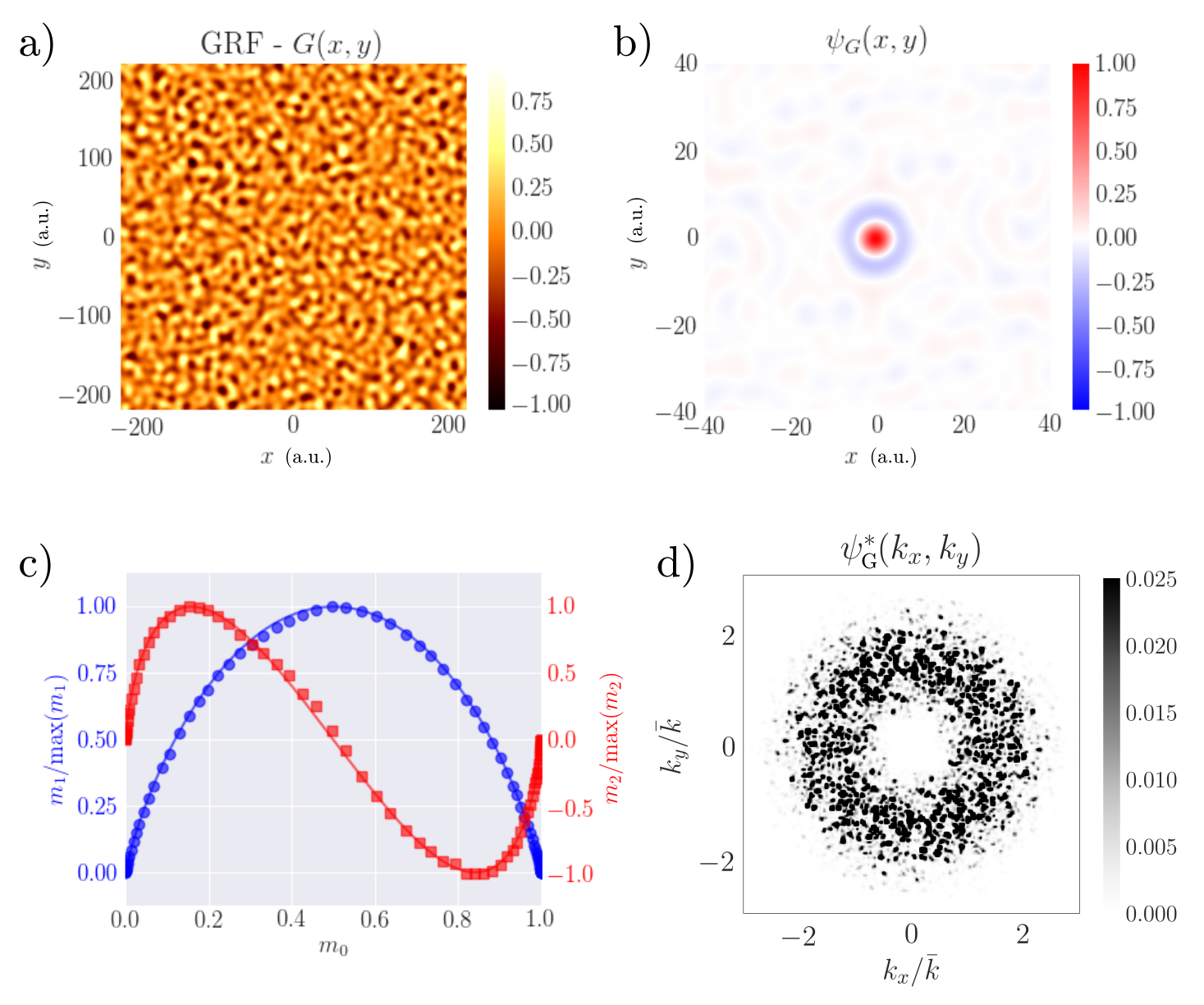}   
    \caption{a) Numerically generated GRF. b) Auto-correlation function of a). c) Minkowski functionals analysis of a). d) Spectral density of b).}
    \label{fig:SIfigure6}
\end{figure}

\section{Pattern Analysis}

The analysis of patterns through Minkowski functionals follows the work reported in Ref.~\cite{SIMantz_2008}. The Minkowski functionals, $M_i(A)$ with $A$ a compact 2D domain, read
\begin{equation}
    \begin{split}
        M_0(A)&=\int_A d\mathbf{x}\\
        M_1(A)&=\frac{1}{2\pi} \int_{\partial A} d\mathbf{x}\\
        M_2(A)&=\frac{1}{2\pi^2} \int_{\partial A} \frac{1}{R}d\mathbf{x}
    \end{split}
\end{equation}
with $d^n\mathbf{x}$ the Lebesgue measure of n-dimensional Euclidean space on which the integration in performed, and $R$ the radius of curvature of $\partial A$. These functionals contain the following information: $M_0(A)=|A|$ with $|A|$ the area of $A$, $M_1(A)=(1/2\pi)U$ with $U$ the boundary length of $A$ and $m_1(A)=(1/\pi)\chi$ with $\chi$ its Euler characteristic.

Here, we use the formulation discretized and averaged Minkowski functionals together with
the marching cube/marching square algorithm for grey-scale images implemented in the software enclosed with Ref.~\cite{SIMantz_2008}.
In particular we consider 8-bit gray-scale images representing the thickness of samples by a space dependent field $\rho(\mathbf{x})\in[0,255]$. Then we evaluate the averaged Minkowski functionals
\begin{equation}
 m_i(\bar{\rho})=\frac{1}{|\Omega|}M_i(\mathcal{B}_{\bar{\rho}})
\end{equation}
with $|\Omega|$ the total size of the image and $\bar{\rho}$ a given threshold defining a binary image 
\begin{equation}
\mathcal{B}_{\bar{\rho}}=\Theta(\rho(x)-\bar{\rho}),
\end{equation}
with $\Theta$ the Heavyside function. Thus, $m_0(\bar{\rho})$ corresponds to the fraction of $|\Omega|$ occupied by the non-zero region in $\mathcal{B}_{\bar{\rho}}$. $m_1(\rho)$ represent the normalized boundary length $U(\bar{\rho})$ between the regions $\mathcal{B}_{\bar{\rho}}=1$ and $\mathcal{B}_{\bar{\rho}}=0$ and $m_2(\rho)$ corresponds to the averaged Euler characteristic $\chi(\bar{\rho})$. 
Considering $m_{1,2}(m_0)$ with the encoded averaging technique allows for obtaining results that are independent of image conditions such as saturation or contrast \cite{SIMantz_2008}. 

\begin{figure}[b]
    \centering
    \includegraphics[width=\linewidth]{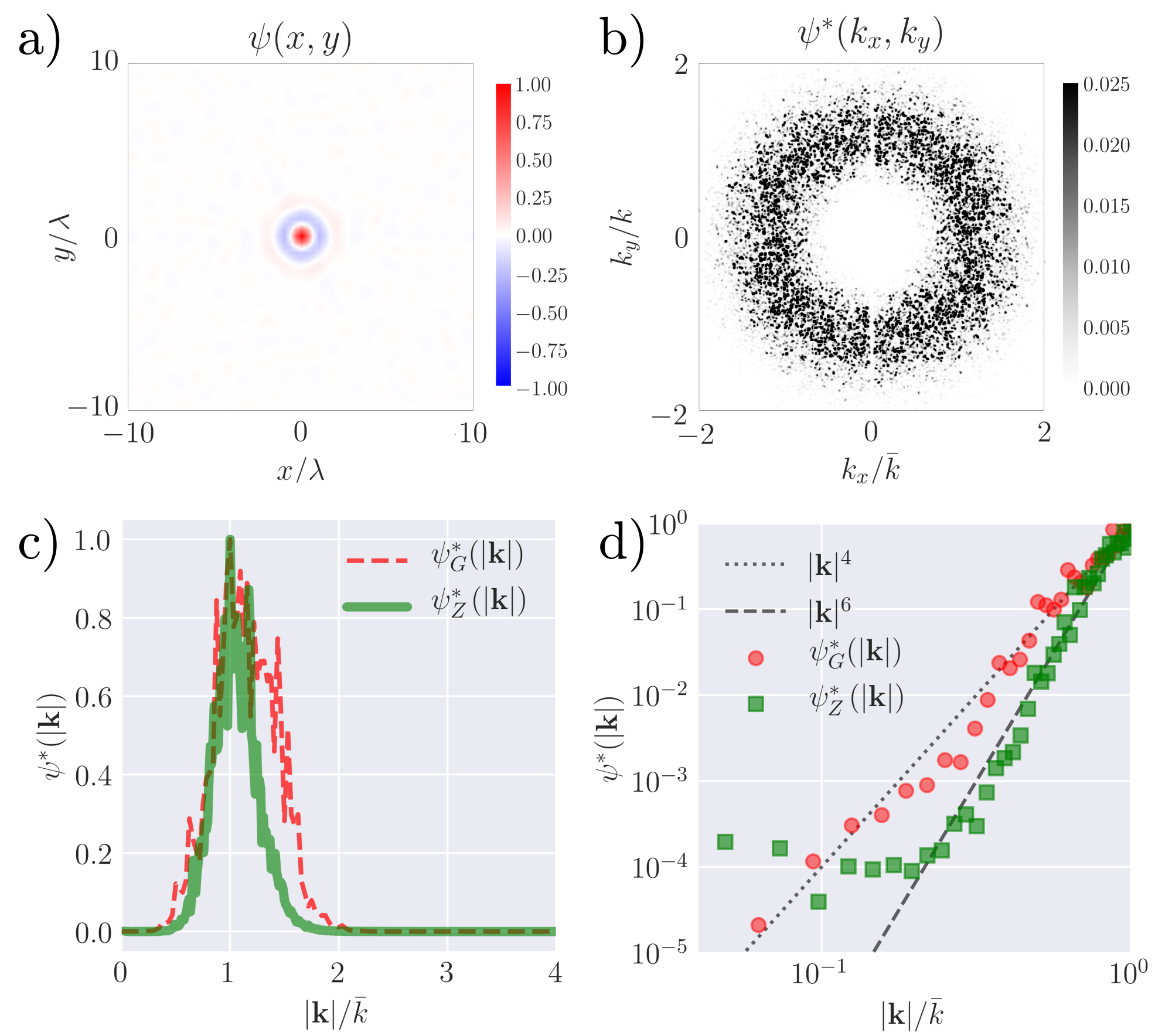}   
    \caption{\textit{Analysis of hyperuniformity}: (a) Spatial auto-correlation function $\psi(x/\lambda,y/\lambda)$, with $\lambda$ the characteristic length of $\psi$. (b) spectral density $\psi^*(k_x,k_y)$ and (c) $\psi^*(|\mathbf{k}|)$ (details in text) are shown for the experimental pattern analyzed in Fig.~2(c) in the main text. (d) Logarithmic plot highlighting the decay of $\psi^*(|\mathbf{k}|)$ $ |\mathbf{k}|\rightarrow 0$. Panels (c) and (d) show $\psi^*(|\mathbf{k}|)$ for both the experimental height profile ($\psi_{Z}^*(|\mathbf{k}|)$) and a numerically generated Gaussian random field  ($\psi_{G}^*(|\mathbf{k}|)$) as in Fig.~\ref{fig:SIfigure6}.}
    \label{fig:figure3SSII}
\end{figure}

In the main text, the analysis of patterns emerging from experiments and simulations is compared to the Minkowski functionals obtained in the case of a Gaussian Random Field (GRF). This corresponds to a field with Gaussian probability density functions which can be obtained by sum of a large number of plane waves with uniformly distributed phases. The averaged Minkowski functionals for a GRF are given by the following expressions (with $\rho$ a value in the grey scale for pixel) \cite{SIMantz_2008}.
\begin{equation}
    \begin{split}
        m_0^{\rm GRF}(A)&=\frac{1}{2}\left[1-\text{erf}\left(\frac{\rho-\rho_0}{\sqrt{2}\sigma}\right) \right]\\
        m_1^{\rm GRF}(A)&=\frac{k}{\sqrt{8\pi}}e^{-\frac{1}{2}(\rho-\rho_0)/\sigma)^2}\\
        m_2^{\rm GRF}(A)&=\frac{k^2}{\sqrt{2\pi^3\sigma^2}}(\rho-\rho_0)e^{-\frac{1}{2}(\rho-\rho_0)/\sigma)^2}\\
    \end{split}
\end{equation}
Following \cite{SIMantz_2008} we set $k=1$, $\sigma=1$, $\rho_0=0$. A numerically generated GRF and its analysis are shown in Fig.~\ref{fig:SIfigure6}.

The analysis of the hyperuniform nature of the patterns emerging by the spinodal solid-state dewetting process has been performed following Refs.~\cite{SITorquato2016,SIMa2017,SIGruhn_2016}. The height profile of the structures, $H(x,y)$, is considered by its auto-correlation function $\psi_H(x,y)$. $\psi_H^*(k_x,k_y)$ denotes the 2D Fourier transform of $\psi_H(x,y)$ with $k_x,k_y$ the components of the (2D) wave vectors. $\psi(r)$ and $\psi^*(|k|)$ denote the same quantities evaluated as radial distributions in the corresponding domains, with $r=\sqrt{x^2+y^2}$ and $|k|=\sqrt{k_x^2+k_y^2}$ the distance from the center of the frame of reference in the real and reciprocal space, respectively. Length scales are scaled according to the characteristic length emerging in the pattern. This consists of $\bar{k}$ and $\lambda=2\pi/\bar{k}$ in the reciprocal and real space, respectively. With this choice, the quantities are independent of image resolution and size, provided that these are large enough to resolve $\lambda$. To assess the hyperuniformity character of the patterns the decay of $\psi^*(|k|)$ for $\psi^*(|k|\rightarrow 0)$ is evaluated in detail. Patterns can be considered to be hyperuniform if $\psi^*(|k|) \leq |k|^{4}$ and $\psi^*(|k|)/\psi^*(\bar{k}) \leq 10^{-4}$ for $|k| \rightarrow 0$ \cite{SITorquato2016,SIMa2017}.

We report here further on the analysis entering Fig.~3(a) in the main text. Fig.~\ref{fig:figure3SSII}(a) shows explicitly the correlation function $\psi(x,y)$ of the height profile $Z(x,y)$ extracted from of Fig.~\ref{fig:SIfigure3}. Then, its spectral density $\psi^*$ is shown by means of different representations, namely $\psi^*(k_x,k_y)$ (Fig.~\ref{fig:figure3SSII}(b)), $\psi^*(|\mathbf{k}|)$ (Fig.~\ref{fig:figure3SSII}(c)), and a focus on $\psi^*(|\mathbf{k}|)$ for $|\mathbf{k}|\rightarrow 0$ in a logarithmic plot (Fig.~\ref{fig:figure3SSII}(d).


\begin{thebibliography}{64}
\bibitem{FOOTNOTE}
  {Results in terms of wavelength and emerging patterns are
  independent on the initial perturbation, provided that it contains
  wavelengths smaller than $\ell $. For wavelengths larger than $\ell $,
  stripes are obtained (see S.M.)}
\end{thebibliography}
\end{document}